\documentclass[12pt]{article}
\usepackage{graphicx}
\usepackage{bm}
\usepackage{amsmath}
\usepackage{amsfonts}
\usepackage{amssymb}
\usepackage{dsfont}
\usepackage{epstopdf}
\usepackage{epsfig}
\usepackage{setspace}
\usepackage{url}
\usepackage{color}
\usepackage[usenames,dvipsnames,svgnames,table]{xcolor}
\usepackage[authoryear,round]{natbib}
\usepackage{rotating}
\usepackage{geometry}
\usepackage{pdflscape}
\usepackage{longtable, booktabs, tabularx}
\usepackage{caption}
\usepackage{hyperref}
\usepackage{multicol}
\usepackage{listings}
\hypersetup{
  colorlinks   = true, %Colours links instead of ugly boxes
  urlcolor     = blue, %Colour for external hyperlinks
  linkcolor    = blue, %Colour of internal links
  citecolor   = blue, %Colour of citations
}
\newtheorem{prediction}{Prediction}
\newtheorem{result}{Result}

\begin{document}

\title{Contests with sequential moves:\\ An experimental study}
\author{Arthur B. Nelson\thanks{Florida Department of Transportation, Tallahassee, FL 32399, USA, \url{NelsonArthur53@gmail.com}.} \and Dmitry Ryvkin\thanks{School of Economics, Finance and Marketing, RMIT University, Melbourne, VIC 3000, Australia, \url{d.ryvkin@gmail.com}.}}
\date{\today}

\maketitle

\begin{abstract}
\noindent We study experimentally contests in which players make investment decisions sequentially, and information on prior investments is revealed between stages. Using a between-subject design, we consider all possible sequences in contests of three players and test two major comparative statics of the subgame-perfect Nash equilibrium: The positive effect of the number of stages on aggregate investment and earlier mover advantage. The former prediction is decidedly rejected, as we observe a reduction in aggregate investment when more sequential information disclosure stages are added to the contest. The evidence on earlier mover advantage is mixed but mostly does not support theory as well. Both predictions rely critically on large preemptive investment by first movers and accommodation by later movers, which does not materialize. Instead, later movers respond aggressively, and reciprocally, to first movers' investments, while first movers learn to accommodate those responses.\\



\end{abstract}

\newpage

\onehalfspacing

\section{Introduction}
\label{sec:ontro}

In \textit{contests}, agents spend resources attempting to secure a valuable prize. Examples include R\&D competition, lobbying, political campaigns, competition for promotion or bonuses in organizations, and sports. Many contests are \textit{sequential}, with some agents making investment decisions later than others. In this case, information about prior investments can play a crucial role. 

In this paper, we study experimentally sequential contests in which players make investments in stages, and after each stage prior investments are revealed. \cite{Hinnosaar:2024} has proposed a novel approach to the subgame-perfect Nash equilibrium (SPNE) analysis of sequential contests that yields two major predictions. First, aggregate investment should increase with the granularity of information revealed. That is, for a given total number of participants, introducing more sequential stages leads to an increase in aggregate investment. Second, there is a universal earlier-mover advantage: In any sequential contest with three or more participants earlier movers invest more, win more often and earn higher payoffs than later movers. We test these predictions using a laboratory experiment. 

In the experiment, we utilize a between-subject design to compare behavior in all possible sequential contests of three players. There are four treatments---(3), (1,2), (2,1) and (1,1,1)--where the dimension of each vector $(n_1,\ldots,n_T)$ is the number of stages, and components $n_t$ give the number of players making investment decisions at stage $t$. Treatment (3) is the standard simultaneous-move contest, (1,1,1) is fully sequential, and the other two treatments are in between. We use the lottery contest success function (CSF) of \cite{Tullock1980} to determine the probability of each player winning given the final vector of investments. 

In the SPNE, aggregate investment increases in the number of stages, and earlier-mover advantage is predicted in all sequential treatments. These predictions rely heavily on behavior being consistent with backward induction; yet, it is well documented that human subjects are generally bad at backward induction, especially in longer games  \citep[e.g.,][]{Mckelvey-Palfrey:1992,Fey-et-al:1996,Binmore-et-al:2002}, although hyper-rational subjects may be very good at it \citep[e.g.,][]{Palacios-Huerta-Volij:2009}. It is, therefore, reasonable to expect that at least some parts of the SPNE predictions will fail in our experiment, especially because the contest game is arguably more complex than the centipede or alternating offer bargaining games those earlier studies are based on. However, it is not clear \textit{a priori} how exactly the SPNE predictions may fail. It is well-known, for example, that, on average, subjects tend to invest substantially above equilibrium in contest experiments \citep[e.g.,][]{Sheremeta2013}. Will such overbidding continue in sequential contests? Will earlier movers or later movers overbid even more, or will some of them perhaps underbid? What consequences (if any) will those deviations from the SPNE have for aggregate effort in various sequential scenarios? A controlled laboratory experiment allows us to answer these questions and explore the underlying behavioral mechanisms.

The idea that earlier movers have a strategic advantage in markets goes back at least to Stackelberg who studied sequential duopoly with linear demand.\footnote{\cite{Stackelberg:2010} is the first English translation of the original book published in German in 1934.} However, due to a peculiar shape of best responses, there is no impact of sequential moves on equilibrium investment in two-stage Tullock contests of two players \citep{Linster}.\footnote{The first-mover advantage arises in two-stage contests of more than two players \citep{DixitContest}.} Thus, three is the minimum number of players for which sequential moves produce any differences in equilibrium. Exploring multi-stage contests with more than two players is technically challenging, and until recently only results for several special cases were available \citep{Glazer2000,Kahana-Klunover:2018,DixitContest}. The inverse best response approach of \cite{Hinnosaar:2024} allows for a systematic analysis of a wide class of arbitrary multi-stage games, including contests.

While the experimental literature on behavior in contests is vast \citep[for a review see, e.g.,][]{Dechenaux2015}, the literature on sequential contests is rather scarce. \cite{FONSECA2009582} compared behavior in two-player sequential and simultaneous-move contests, focusing primarily on the effect of heterogeneity in players' abilities. In his baseline treatments with symmetric players, \cite{FONSECA2009582} observed behavior consistent with the neutrality prediction of \cite{Linster}, i.e., no differences in investment between simultaneous and sequential contests. \cite{Nelson:2020} studied entry deterrence in two-player and three-player sequential contests. The goal of that study was to test the prediction of \cite{Linster} and its breakdown when transitioning from two to three players. However, motivated by third-party entry in political competition with two stable incumbent parties, \cite{Nelson:2020} only considered one type of three-player sequential contests---those involving two first movers.\footnote{Sequence (2,1), using the taxonomy of treatments in this paper.} In the present paper, we reuse data from two three-player treatments of \cite{Nelson:2020}---simultaneous and (2,1)---and add two other treatments---(1,2) and (1,1,1)---to consider all possible move sequences and directly test the predictions of \cite{Hinnosaar:2024}.

The rest of the paper is structured as follows. Section \ref{sec:theory} summarizes the theoretical model and main predictions, and Section \ref{sec:design} presents the experimental design. Results are reported in Section \ref{sec:results} and discussed in Section \ref{sec:discussion}.

\section{Theory and predictions}
\label{sec:theory}

We consider sequential contests of $n=3$ symmetric, risk-neutral players $i=1,2,3$ choosing investments $x_i\in\mathds{R}_+$. Investments are made in stages, and at the beginning of each stage all prior investments are revealed. The winner of the contest is determined after all three players made their investment decisions. The probability of player $i$ winning is given by the lottery contest success function (CSF) of \cite{Tullock1980},
\begin{equation}
\label{csf}
p_i = \left\{\begin{array}{ll}
\frac{x_i}{\sum_{j=1}^n x_j}, & \text{if } \sum_{j=1}^n x_j>0 \\[3ex]
\frac{1}{n}, & \text{if } \sum_{j=1}^n x_j=0\\
\end{array}\right. 
\end{equation}
The winner receives a prize $V>0$, and all players lose their investments. 

A sequence of moves in a contest of $n$ symmetric players is completely characterized by a $T$-dimensional vector ${\bf n}=(n_1,\ldots,n_T)$, where $T\ge 1$ is the number of stages, $n_t\ge 1$ is the number of players making investment decisions at stage $t$, and $\sum_{t=1}^Tn_t=n$. For $n=3$, there are four possible sequences: (3) is a one-stage simultaneous-move contest, (1,2) and (2,1) are two-stage Stackelberg-like contests with one leader and two followers, and with two leaders and one follower, respectively; finally, (1,1,1) is a three-stage contest where one player makes a decision at each stage.

The solution concept is subgame-perfect Nash equilibrium (SPNE). Let $X$ denote aggregate equilibrium investment in a sequential contest ${\bf n}$ with unit prize. Following the inverse best response method developed by \cite{Hinnosaar:2024}, $X$ can be calculated as the largest root of the equation $f_0(X)=0$, where functions $f_0,f_1,\ldots,f_T$ are obtained recursively using the relation $f_{t-1}(X) = f_t(X)-n_tf'_t(X)X(1-X)$, with terminal condition $f_T(X)=X$. Individual equilibrium investment of players making decisions at stage $t$ can then be found as $x_i^* = \frac{1}{n_t}[f_t(X)-f_{t-1}(X)]$. In the experiment, we use $n=3$ and prize $V=240$. The resulting individual and aggregate equilibrium investment is shown in Table \ref{tab:summary}. We follow the convention that player $i$ (weakly) precedes player $j$ for $i<j$. Based on Table \ref{tab:summary}, there are two main predictions: the first one describes the ranking of aggregate investment across treatments, and the second characterizes the preemptive behavior of early movers.

\begin{prediction}
\label{h1}
Let $X_{k}$ denote the observed aggregate investment in treatment $k$. Then $X_{(3)}<X_{(1,2)}=X_{(2,1)}<X_{(1,1,1)}$.
\end{prediction}

\begin{prediction}
\label{h2}
Earlier movers in treatments (1,2), (2,1) and (1,1,1) invest more than later movers, and more than players in (3).
\end{prediction}

\section{Experimental design}
\label{sec:design}

\paragraph{Preliminaries} The experiment followed a between-subject design with four treatments corresponding to the four possible move sequences in three-player contests. A total of 333 subjects (64\% of them female) were recruited using ORSEE \citep{Greiner2015} from the population of $\sim$3000 students at Florida State University who preregistered for participation in social science experiments. There were 81, 90, 81 and 81 subjects in treatments (3), (1,2), (2,1) and (1,1,1), respectively. A total of 21 sessions with 9 or 18 subjects each were conducted in the XSFS Lab at FSU, with subjects making decisions at visually separated computer terminals. The experiment was implemented in zTree \citep{Fischbacher2007}. On average, sessions lasted $\sim$60 minutes, and subjects earned \$21.17 including a \$7 participation payment.

\paragraph{Procedures}  Each session consisted of three parts. Instructions for each part were distributed on paper at the beginning of that part and read out loud by the experimenter (sample instructions are available in Online Appendix \ref{app_inst}). In Part 1, subjects' risk attitudes were assessed using the method of \cite{HoltLaury}. Subjects made 10 choices between two lotteries, $A=(\$2.00,\$1.60;p,1-p)$ and $B=(\$3.85,\$0.10;p,1-p)$, with $p$ taking values $0.1,0.2,\ldots,1.0$. The outcomes of this part were withheld until the end of the session when one of the 10 choices was selected randomly and played out for actual earnings.

In Part 2---the main part of the experiment---subjects were randomly divided into fixed matching groups of 9 and only interacted within those groups. Subjects went through 25 identical rounds of the contest game with moves sequenced, and information revealed across stages, according to the treatment. Roles were randomly assigned and fixed. At the beginning of each round, subjects within each matching group were randomly matched into groups of three. Each subject was given an endowment of 240 points and could invest any integer number of points between 0 and 240 into the contest. After all subjects made their investment decisions, one winner within each group was randomly determined according to CSF (\ref{csf}). The winner received a prize of 240 points, and all subjects lost their investments.  One round of the 25 was chosen at the end of the session to base subjects' actual earnings on, at the exchange rate of 20 points = \$1. 

In Part 3, we administered a short questionnaire. Subjects reported their gender,  age, major, and self-assessed competitiveness measured on a Likert scale from 1 (``much less competitive than average'') to 5 (``much more competitive than average''). After Part 3, earnings from all parts were calculated and revealed, and  subjects were paid privately by check.

\section{Results}
\label{sec:results}

\begin{table}[tbp]
\centering
{\small 
\begin{tabular}{lcccccccc}
\hline\hline
& \multicolumn{8}{c}{Treatments} \\
& \multicolumn{2}{c}{(3)} & \multicolumn{2}{c}{(1,2)} & \multicolumn{2}{c}{(2,1)} & \multicolumn{2}{c}{(1,1,1)}\\
& SPNE & Observed & SPNE & Observed & SPNE & Observed & SPNE & Observed \\
\hline
$x_1$ 	&	53.33 	&84.85 	& 90 & 88.98	&	67.5& 75.73 &	86.18 & 64.42 \\
		&       	&(6.29) &    & (8.89)	&       & (7.24)&         & (6.35)\\
$x_2$ 	&	53.33 	&84.85 	& 45 & 70.84	& 	67.5& 75.73 & 	63.10 & 62.40 \\
		&       	&(6.29) &    & (3.91)	&       & (7.24)&         & (7.72)\\
$x_3$ 	&	53.33 	&84.85 	& 45 & 70.84	& 	45 	& 68.30 & 	40.01 & 79.66 \\
		&       	&(6.29)	&    & (3.91)	&       & (7.39)&         & (6.12)\\
$X$	  	& 160   	&254.54 &180 & 230.65	& 180	& 219.76& 189.29  & 206.48\\
		&       	&(18.88)&    & (13.96)	&       & (13.50)&        & (15.32)\\
\hline
Clusters&       	&9		&    &	10		&       & 9		&        & 	9	\\
Rounds	&       	&25		&    &	25		&       & 25	&        & 25	\\
$N$		&        	&2,025	& 	 &	2,250	&       & 2,025	&        & 2,025\\
\hline\hline
\end{tabular}
}
\caption{SPNE predictions and summary statistics (averages using data from all rounds). Standard errors clustered by matching group in parentheses.}
\label{tab:summary}
\end{table}

Table \ref{tab:summary} shows average individual and aggregate investments by treatment, along with the SPNE predictions. Individual investment $x_i$ is indexed by $i=1,2,3$, where player $i$ makes a decision no later than player $j$ for $i<j$. In what follows, we employ both parametric and nonparametric tests, with clustering at the matching group level in the former and matching group as the unit of observation in the latter. Thus, unless stated otherwise, the number of independent observations is either 9 or 10 for within-treatment comparisons and 18 or 19 for comparisons between any two treatments. All tests are two-sided.\footnote{While most of our predictions are directional, we adopt two-sided tests to be able to reject a null hypothesis also when the corresponding prediction is reversed.}

\subsection{Aggregate investment}

\begin{figure}[t]
\begin{center}
\includegraphics[width=4.5in]{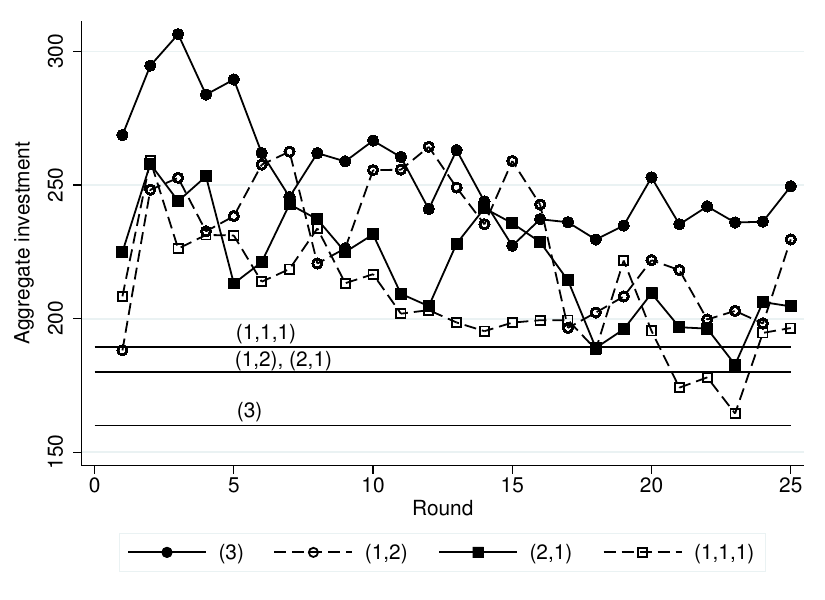}
\end{center}
\caption{Average aggregate investment as a function of time, by treatment. The horizontal lines show the SPNE predictions.}
\label{fig:X}
\end{figure}

We start with the analysis of aggregate investment, $X$. As seen from Table \ref{tab:summary}, the observed averages over all rounds exceed the SPNE predictions. The difference is statistically significant in all treatments except (1,1,1) ($p=0.001$, $0.006$, $0.019$ and $0.295$ for (3), (1,2), (2,1) and (1,1,1), respectively, the Wald test). Such excessive investment (often termed ``overbidding'') is rather standard for contest experiments \citep[see, e.g., a review by][]{Sheremeta2013}. Figure \ref{fig:X}, showing average aggregate investment by round, indicates that overbidding declines over time. A negative time trend is significant in linear regressions of investment on $Round$ in all but one treatment ($p=0.012$, $0.170$, $0.048$ and $0.018$, same order as above, the Wald test). As a result, (3) is the only treatment where statistically significant aggregate overbidding is sustained in the last five rounds of the experiment ($p=0.001$, $0.118$, $0.196$ and $0.643$, same order as above, the Wald test).

More importantly, Table \ref{tab:summary} and Figure \ref{fig:X} show that aggregate investment appears to \textit{decline} in the number of stages. The decline in investment from one treatment to the next is not large enough to register statistically significant differences between adjacent treatments pairwise,\footnote{Pairwise, there is a decline from (3) to (2,1) and from (3) to (1,1,1) in later rounds, but no other comparisons are statistically significant.} but the overall decline is significant, especially in later rounds, as evidenced by the Jonckheere–Terpstra test for trend across the four treatments ($p=0.071$ for all rounds, 0.036 for the last 10 rounds, and 0.011 for the last 5 rounds; $N=37$ and 4 groups in all cases).

\begin{result}
\label{res_X} Aggregate investment decreases in the number of stages, more so in later rounds.
\end{result}
Result \ref{res_X} is in striking contradiction to Prediction \ref{h1} stating that aggregate investment should increase with the number of stages. We look at individual investment for clues on why we observe this pattern. Given the presence of substantial dynamics in the data (cf. Figure \ref{fig:X}), in what follows we restrict attention to the last five rounds to focus on the converged behavior of experienced subjects.

\subsection{Individual investment}

\begin{figure}[t]
\begin{center}
\includegraphics[width=4.5in]{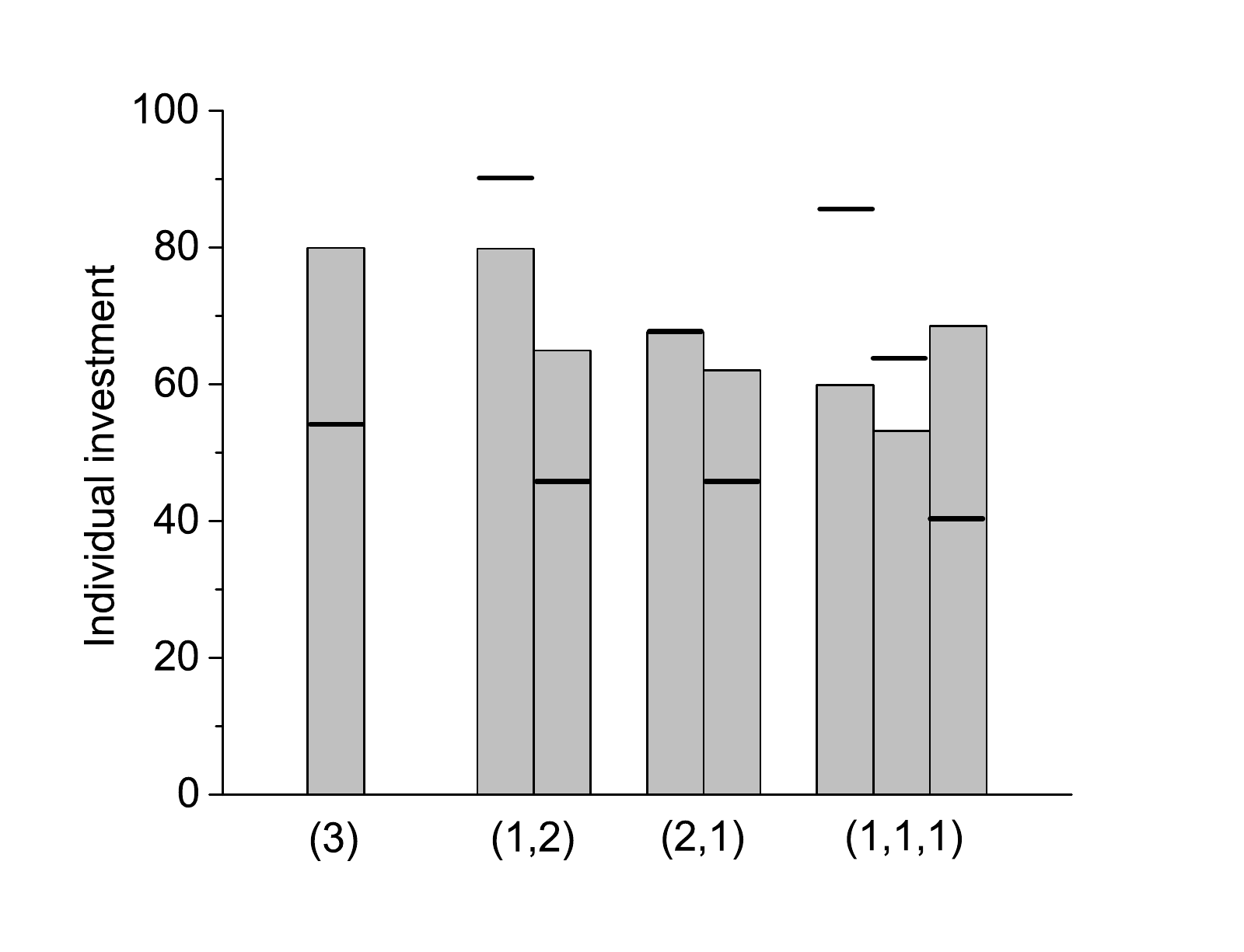}
\end{center}
\caption{Average individual investment, by treatment and move order, using data from the last five rounds of the experiment. The horizontal black lines show the SPNE predictions.}
\label{fig:x_individual}
\end{figure}

Average individual investment by treatment and role, using data from the last five  rounds, is shown in Figure \ref{fig:x_individual}. First, we compare individual investment to the SPNE predictions.\footnote{We also ran regressions  to uncover associations of subjects' behavior with individual characteristics such as gender, risk aversion, self-assessed competitiveness and field of study. These regressions did not produce any systematic results. One notable regularity is the negative effect of risk aversion on investment, but it is only statistically significant for subjects in (3) and first and third movers in (1,1,1). It could be expected that risk aversion plays a role especially in first movers' decisions in the sequential treatments because they are the most exposed in terms of strategic uncertainty; yet, we do not observe this pattern. The results are available in Online Appendix \ref{app_dem}.} In the simultaneous treatment (3), there is significant overbidding of about 50\% ($p<0.001$, the Wald test), consistent with most other contest experiments. In contrast, in the sequential treatments whether or not overbidding is present depends on the role. In (1,2) and (2,1), first movers' investment is in line with the SPNE ($p=0.392$ and 0.984, respectively, the Wald test), while second movers overbid significantly ($p=0.005$ and 0.045, respectively, the Wald test). In (1,1,1), first movers \textit{underbid} ($p=0.027$, the Wald test), second movers are in line with the SPNE ($p=0.206$, the Wald test), and third movers overbid ($p=0.002$, the Wald test).

Comparing investment by players in different roles within each of the sequential treatments, we find no statistically significant differences between the bids of first and second movers ($p=0.226$, 0.638 and 0.549 in (1,2), (2,1) and (1,1,1), respectively, the Wald test). Third movers in (1,1,1) invest similar to first movers ($p=0.463$, the Wald test) and above second movers (this difference is only marginally significant, $p=0.085$, the Wald test). We also compare average individual investment in (3) to the investment of early movers in the sequential treatments. There are no statistically significant differences between investment in (3) and first movers' investments in (1,2) and (2,1) ($p=0.991$ and 0.138, respectively, the Wald test). However, both first and second movers in (1,1,1) invest less than subjects in (3), although the latter difference is only marginally significant ($p=0.006$ and 0.075, respectively, the Wald test).

\begin{result}
\label{res_ind1}
(a) First movers and second movers invest similarly in the sequential treatments. Third movers in (1,1,1) invest more than second movers.

(b) First movers in (1,2) and (2,1) invest at the same level as subjects in (3); first and second movers in (1,1,1) invest less than subjects in (3).
\end{result}
Result \ref{res_ind1} provides no support for Prediction \ref{h2} stipulating preemptive behavior by early movers and accommodation by later movers in the sequential treatments. Instead, first movers in these treatments bid at the level of subjects in (3) or less, while later movers tend to match the investment of early movers. We will now present \textit{ex post} analysis exploring potential explanations for this behavior.

\subsubsection{Joy of winning}
\label{sec:jow}

Competitive preferences, such as joy of winning, received considerable attention in the contest literature and have been used to explain many experimental findings \citep[e.g.,][]{Sheremeta2013,Rockenbach-Waligora:2016}. For example, overinvestment in (3) is consistent with a basic joy of winning model positing that in addition to the monetary prize $V$ there is a nonmonetary reward for winning $w$, so that the total effective prize in the contest is $V+w$ \citep[e.g.,][]{Goeree-et-al:2002}. Using average individual investment in (3) in the last five rounds, 79.94, we estimate $w=119.73$, about 50\% of the monetary prize.\footnote{This estimation is based on matching the observed average investment in (3) to the symmetric NE with joy of winning, $x^{\ast,w}_{(3)}=\frac{2(V+w)}{9}$.} Assuming that the same joy of winning is present, on average, in the sequential treatments,\footnote{It is, of course, possible that subjects enjoy winning differently depending on their order of moves in the sequential treatments; for example, it may be more fun to win moving last. While role-dependent joy of winning could fit the data really well, it would be too cheap an explanation, and unlikely to be the whole story. A single-parameter model calibrated out of sample strikes us as a more prudent approach.} we derive the corresponding SPNE predictions adjusted for joy of winning. These, along with average individual and aggregate investments in the last five rounds, are presented in Table \ref{tab:SPNE_w}.

\begin{table}[tbp]
\centering
{\small 
\begin{tabular}{lcccccc}
\hline\hline
& \multicolumn{6}{c}{Treatments} \\
& \multicolumn{2}{c}{(1,2)} & \multicolumn{2}{c}{(2,1)} & \multicolumn{2}{c}{(1,1,1)}\\
& SPNE($w$) & Observed & SPNE($w$) & Observed & SPNE($w$) & Observed \\
\hline
$x_1$ 	& 134.90 & 79.81&	101.17& 67.64 &	129.17 & 59.86 \\
		&    & (11.34)	&       & (6.55)&         & (9.76)\\
$x_2$ 	& 67.45 & 64.95	& 	101.17& 67.64 & 	94.58 & 53.19 \\
		&    & (5.34)	&       & (6.55)&         & (7.20)\\
$x_3$ 	& 67.45 & 64.95	& 	67.45	& 62.04 & 	59.97 & 68.51 \\
		&    & (5.34)	&       & (7.18)&         & (6.19)\\
$X$	  	&269.80 & 209.71& 269.80& 197.31& 283.72  & 181.56\\
		&    & (17.20)	&       & (12.27)&        & (16.04)\\
\hline
Clusters&    &	10		&       & 9		&        & 	9	\\
Rounds	&    &	5		&       & 5		&        & 5	\\
$N$		& 	 &	450		&       & 405	&        & 405  \\
\hline\hline
\end{tabular}
}
\caption{SPNE predictions adjusted for joy of winning calibrated from (3) and average investments using data from the last five rounds. Standard errors clustered by matching group in parentheses.}
\label{tab:SPNE_w}
\end{table}

Since the joy of winning correction is simply a multiplier on the SPNE predictions from Table \ref{tab:summary}, the predicted comparative statics across treatments and roles are preserved. It is, however, instructive to look at the point predictions. As seen from Table \ref{tab:SPNE_w}, aggregate investment in the sequential treatments falls short of the adjusted predictions ($p<0.01$ in all cases). This is due to significant underinvestment by early movers ($p<0.01$ for first movers in all the treatments and for second movers in (1,1,1)). In contrast, second movers in (1,2) and (2,1), and third movers in (1,1,1) invest in line with the adjusted SPNE ($p=0.651$, 0.473 and 0.205, respectively).

\subsubsection{Preemptive investment by first movers}
\label{sec:br}

The results in Table \ref{tab:SPNE_w} suggest that underinvestment by early movers is the main source of the reversal of the SPNE comparative statics. One potential explanation is that early movers \textit{fail to preempt}; that is, they do not optimally incorporate possible responses by later movers into their investment decisions. To explore this further, we estimate later movers' response functions of the form $r_2=\beta_0+\beta_1m_1+\gamma_1m_1^2+u_2$ for second movers and $r_3=\beta_0+\beta_1m_1+\gamma_1m_1^2+\beta_2m_2+\gamma_2m_2^2+u_3$ for third movers.\footnote{We include the quadratic terms to capture the nonmonotonicity of best response functions in Tullock contests. While \cite{Rockenbach-Waligora:2016} find that linear (increasing) response functions describe behavior well in their experiment, we have evidence of nonmonotonicity in ours.} Here, $r_2$ and $r_3$ are the later movers' investments, $m_1$ and $m_2$ are the investments by earlier movers those later movers observe in the corresponding sequential treatments, and $u_2$ and $u_3$ are idiosyncratic errors.\footnote{Because there are two first movers in (2,1), we define $m_1$ to be the average of their investments in that treatment.} Table \ref{tab:br_est} presents the results of pooled OLS regressions estimating the response functions using data from the first 20 rounds.

\begin{table}[tbp]
\centering
{\small 
\begin{tabular}{lcccc}
\hline\hline
& \multicolumn{4}{c}{Treatments} \\
& (1,2) & (2,1) & \multicolumn{2}{c}{(1,1,1)}\\
& $r_2$ & $r_2$ & $r_2$ & $r_3$ \\
\hline
$m_1$ 	& 0.091	& 0.249	&0.103	& 0.020 \\
		&(0.107)&(0.190)&(0.243)& (0.131)\\
$m_1^2$	& 9.6$\times 10^{-5}$& -0.0020*&-7.1$\times 10^{-4}$& -1.2$\times 10^{-4}$ \\
		&(4.4$\times 10^{-4}$)&(0.0010)&(9.3$\times 10^{-4}$)& (6.7$\times 10^{-4}$)\\
$m_2$ 	&  		& 		& 		& 0.333 \\
		&    	& 		&       & (0.189)\\
$m_2^2$	& 		&		&		& -8.1$\times 10^{-4}$ \\
		&		&		&		& (8.0$\times 10^{-4}$)\\
Intercept&62.72*** & 67.60***	& 	63.93***	& 66.76*** \\
		& (5.85)& (5.04)& (10.97)& (10.91)\\
\hline
Clusters& 10   	& 9		& 9     & 9\\
Rounds	& 20   	& 20	& 20    & 20\\
$N$	  	&1,200 	& 540	& 540	& 540\\
$R^2$	& 0.015	& 0.014	& 0.0065 & 0.025\\
\hline\hline
\end{tabular}
}
\caption{Pooled OLS regression results for linear-quadratic response models by later movers using data from the first 20 rounds, by treatment. Standard errors clustered by matching group in parentheses. Significance levels: * $p<0.1$, ** $p<0.05$, *** $p<0.01$.}
\label{tab:br_est}
\end{table}

As seen from Table \ref{tab:br_est}, the linear effects of earlier movers' investment are positive and the quadratic effects are negative in all treatments, with the exception of (1,2) where the quadratic effect is positive but also very small. Almost none of the estimates are statistically significant, indicating a large amount of noise; nevertheless, the presence of negative quadratic effects points at nonmonotone responses, with estimated turning points at $m_1=62.25$ and 71.96 for second movers in (2,1) and (1,1,1), respectively; and at $m_1=83.10$ for third movers in (1,1,1).\footnote{There is no turning point for second movers in (1,2) because the estimated $\gamma_1$ is positive, and no turning point below 240 in $m_2$ for third movers in (1,1,1) because the estimated $\gamma_2$ is too small compared to $\beta_2$.}

With the estimated response functions in hand, we compute the optimal preemptive investment by first movers. The idea is that during the first 20 rounds first movers learn, by observing contest outcomes, how the population of later movers responds to their investment decisions. We also include the joy of winning correction. For example, for treatment (1,2) this computation entails solving
\[
\max_{x_1\in[0,240]} \frac{(240+119.73)x_1}{x_1+2r_2(x_1)} - x_1,
\] 
where it is assumed that the two second movers behave symmetrically, and $r_2(\cdot)$ is the estimated linear-quadratic response function with coefficients from Table \ref{tab:br_est}.\footnote{In (2,1), the computation is a bit more complicated because there are two first movers, and we identify a symmetric equilibrium between them given $r_2(\cdot)$ that is based on their average investment. In (1,1,1), we treat the third mover as responding to $x_1$ by the first mover and $r_2(x_1)$ by the second mover; that is, we assume that the third mover's investment is $r_3(x_1,r_2(x_1))$. \texttt{Mathematica} code for these computations is available in Online Appendix \ref{app_calc}.} The resulting optimal investments for first movers are $x_1=72.03$, 83.11 and 68.48 in (1,2), (2,1) and (1,1,1), respectively.\footnote{The corresponding optimal investments without the joy of winning correction are 48.06, 55.45 and 45.69. We, however, believe that the comparison including the joy of winning correction is more appropriate because our goal is to understand what happens in the sequential treatments as compared to (3), so whatever preferences drive the overinvestment in (3) should be taken into account.} Comparing these to the average investments by first movers in the last five rounds, $x_1=79.81$, 67.64 and 59.86, respectively (see Table \ref{tab:SPNE_w}), we observe that subjects in the role of first movers adjust pretty well to the behavior of later movers. At least, there is no major underinvestment on their part given the later movers' response. Investment is somewhat lower than optimal in (2,1) and (1,1,1), although there is so much noise in the estimates of optimal investments that those differences are not statistically significant.

\subsubsection{Reciprocity and learning/strategic teaching}
\label{sec:other_regarding}

What can explain the behavior of later movers? Despite finding in the previous section that first movers are not that far off anticipating the behavior of later movers, the latter do not invest substantially above the joy of winning-adjusted SPNE, on average (see Table \ref{tab:SPNE_w}). Thus, later movers do not invest that aggressively on average; they do, however, \textit{respond} aggressively to earlier movers' investments. In fact, later movers may be interested in strategically teaching earlier movers to invest less. 

A striking feature of Figure \ref{fig:x_individual}, confirmed by Result \ref{res_ind1}, is that differences in  investments between subjects in different roles in the sequential treatments are less than predicted, if at all present. In other words, contrary to the SPNE predictions, average investments are very similar. Thus, later movers' behavior can be explained by other-regarding preferences, namely, inequality aversion and reciprocity \citep[e.g.,][]{Fehr-Schmidt:1999,Bolton-Ockenfels:2000,Charness-Rabin:2002,Cox-et-al:2007}. These subjects can earn more, in expectation, by investing less than first movers. The gain from such a reduction in investment would, however, be lower for them than for the first movers; hence, sufficiently inequality averse second movers would prefer to equalize investments (and payoffs) instead.\footnote{Investment equalization can also arise from imitation. If a second mover is not sure how much they should invest and views the first mover's investment as ``reasonable,'' she can simply imitate it.} Matching investment can also be interpreted as a form of (negative) reciprocity: If a second mover believes the first mover is too aggressive, she may be willing to reciprocate.\footnote{\cite{Lau-Leung:2010} show that inequality aversion and reciprocity explain followers' behavior in Stackelberg duopoly experiments by \cite{Huck-et-al:2001}.}

\section{Discussion and conclusions}
\label{sec:discussion}

The observed behavior is not consistent with the SPNE (adjusted for joy of winning or not), and the results are essentially the opposite of the comparative statics predictions of \cite{Hinnosaar:2024}: The largest aggregate investment is observed in the simultaneous-move treatment (3), and it declines in the number of stages. Most of these deviations can be explained by a combination of joy of winning and reciprocity.

At first glance, the reduction in investment in the sequential treatments is mostly due to first movers investing substantially below the (joy of winning-adjusted) SPNE predictions. However, our analysis of responses by later movers to first movers' investments reveals that those responses are overly aggressive, teaching the first movers to invest less over time. The resulting investment by first movers in the last five rounds of the experiment is, in fact, consistent with optimization taking the empirical responses of later movers into account (and correcting for joy of winning). The later movers' responses are consistent with reciprocity. A similar cautious behavior by first movers and reciprocal responses by second movers have been observed in Stackelberg duopoly experiments by \cite{Huck-et-al:2001,Huck-et-al:2002} and \cite{Huck-Wallace:2002}, although in those studies aggregate output in Stackelberg markets was still higher than in Cournot (simultaneous-move) markets.

It is, of course, not clear what would have happened had first movers remained more aggressive despite the response they encountered. Later movers could have escalated investment even more, or they could have eventually accommodated.\footnote{Such counterfactuals can be investigated using a strategy method, as in \cite{Huck-Wallace:2002}.} In either case, aggregate investment would likely have increased in the sequential treatments, and the comparative statics of the SPNE would have been preserved. 

To conclude, we argue that, ultimately, the failure to produce aggregate investment in line with the SPNE comparative statics is due to the cautious bidding by first movers that is, in turn, caused by reciprocal behavior of later movers. This study was designed to only test the basic comparative statics of the SPNE, and cannot address the underlying causes too deeply. Further investigation of sequential contests, e.g., studying longer games and carefully controlling for strategic sophistication, competitive preferences and other-regarding preferences, can be of interest.

\bibliographystyle{aea}

\newpage

\appendix

\section*{Supplementary material to ``Contests with sequential moves: An experimental study'' by Arthur B. Nelson and Dmitry Ryvkin}

Section \ref{app_inst} of this online appendix contains experimental instructions for treatment (1,1,1). Section \ref{app_dem} contains the results of regressions of individual investment on subjects' individual characteristics. \texttt{Mathemtica} code for the computations of optimal investment by first movers reported in Section \ref{sec:br} is in Section \ref{app_calc}.

\section{Experimental instructions}
\label{app_inst}

\subsection*{General Information}
Welcome to today's experiment. Please refrain from making noise or communicating with the other participants during the experiment. If you have any questions please raise your hand and someone will come and answer your question privately.

For your participation in this experiment you will receive a show up payment of \$7.00, and will have the opportunity to earn additional money during the experiment. Your payments will depend on the decisions made by you and the other participants. At the end of the experiment you will be paid privately. No other participant will be made aware of your payment.

There will be several stages to this experiment and you will be provided instructions prior to the beginning of each stage.

\subsection*{Part 1}

For your first task today you will be asked to make a choice between a series of lotteries. You will be shown a list of ten pairs of lotteries as shown below:

\begin{figure}[h]
	\centering
	\includegraphics[width=0.7\linewidth]{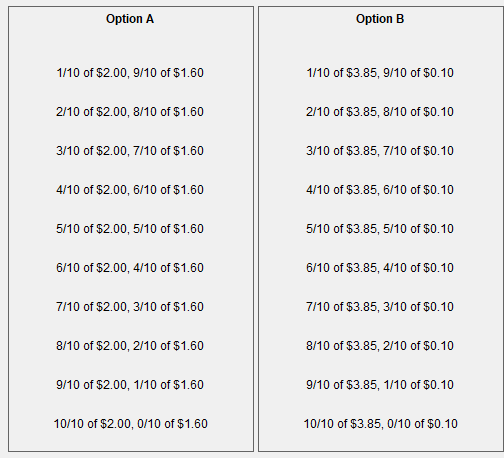}
	\caption{}
	\label{fig:hlcapture}
\end{figure}

Each option has two possible payoffs and an associated probability of getting that payoff. For example in row 2 Option A has a $\frac{2}{10}$ or 20\% chance of paying out \$2.00 and a $\frac{8}{10}$ or 80\% chance of paying out \$1.60. In the same row Option B has a $\frac{2}{10}$ or 20\% chance of paying out \$3.85 and a $\frac{8}{10}$ or 80\% chance of paying out \$0.10.

In row 7 Option A has a $\frac{7}{10}$ or 70\% chance of paying out \$2.00 and a $\frac{3}{10}$ or 30\% chance of paying out \$1.60. In the same row Option B has a $\frac{7}{10}$ or 70\% chance of paying out \$3.85 and a $\frac{3}{10}$ or 30\% chance of paying out \$0.10.

For each pair you will choose either Option A or Option B. After you have made your choices between the lotteries click submit to finalize your decision. Your choices will determine your payout for this part of the experiment via a random process.

Two random numbers between 1 and 10 have been pre-drawn and written on a piece of paper in an envelope that will be placed in the front of the room. At the end of the experiment the envelope will be opened and the numbers will be revealed. The first number determines which of the ten pairs are chosen. The second number determines the outcome. If the second random number is equal to or less than the numerator of the first probability then the first value of your choice will be paid out, if the random number is greater than the numerator then the second value will be paid out.

For example, if the first number is 3 and you chose Option A in the third row you would be in a lottery with a $\frac{3}{10}$ or 30\% chance of paying out \$2.00 and a $\frac{7}{10}$ or 70\% chance of paying out \$1.60. If the second number is a 1, 2, or 3, you would receive \$2.00, if the second number is a 4, 5, 6, 7, 8, 9, or 10 you would receive \$1.60.

\bigskip

Are there any questions before you make your decisions?

\subsection*{Part 2\footnote{These instructions are for treatment (1,1,1). Instructions for other treatments are similar and available from the authors upon request.}}
	All amounts in this portion of the experiment will be expressed in points. The exchange rate will be 100 points =\$5.00 or 1 point= \$0.05. This part of the experiment will have you make decisions over 25 rounds.
	
	\subsubsection*{Endowment and Expenditure}
	At the beginning of this part of the experiment you will be randomly assigned one of three possible roles: Leader, Follower 1 or Follower 2. You will maintain the same role throughout this part of the experiment. In each round you will be randomly matched with two other participants, and given an endowment of 240 points to expend on a contest. Every group will have 1 player of each type. Leader will
	make his or her expenditure decision first, which will then be revealed to Follower 1, who then decides his or her expenditure. Follower 2 then sees the expenditures of Leader and Follower 1 and decides his or her expenditure. You can expend any integer number of points from 0 to 240. You will keep any points
	you choose not to expend. If you win the prize for the round you will receive 240 additional points.
	
	\subsubsection*{Probability of Winning}
	After the expenditure decisions are made the sum of expenditures in your group will be calculated. Then the probability of you winning the prize in that round is given by:
	
	$$\frac{\mathrm{ Your \  Expenditure}}{\mathrm{The \ Sum\ of\ Your\ Group's\ Expenditures}} $$ 
	
	For example, suppose you chose to expend 20 points and another member of your group chose to expend 30 points, and the third member of your group chose to expend 50 points. Then the probability you will win the prize is:
	
	$$ \frac{20}{20+30+50}=\frac{20}{100}=\frac{1}{5}=20\%$$
	
	For another example, suppose you chose to expend 100 points and another member of your group chose to expend 20 points, and the third member of your group chose to expend 40 points. Then the probability you will win the prize is:
	
	$$ \frac{100}{100+20+40}=\frac{100}{160}=\frac{5}{8}=62.5\%$$
	
	\subsubsection*{Payoff in a Given Round}
	After determining the probability that you win the computer will randomly assign you to a player number. The first player will be assigned to the interval from 0 to their probability of winning, player 2 will receive the interval from player 1's probability of winning to player 1's probability of winning plus player 2's probability of winning, player 3 will be assigned the interval from player 1's probability of winning plus player 2's probability of winning to 100. The computer then draws a random number between 0 and 100 to determine which member of your group wins the prize. If the number drawn is in your interval, you will win the prize, otherwise another player in your group will win the prize. Using the first example from above, if the player numbers are assigned in the order of expenditures then player 1 has the interval from 0 to 20, player 2 has the interval from 20 to 50, and player 3 has the interval from 50 to 100. If the number drawn is less than or equal to 20 player 1 wins, if it is larger than 20 but less than 50 player 2 wins, and if it is greater than 50 player 3 wins. \\
	
	The individual payoff is then calculated as follows:
	
	\begin{multicols}{2}		
		If you win: \\
		
		\noindent240 (Endowment)\\
		240 (Prize)\\
		-(Expenditure)\\
		\noindent\makebox[\linewidth]{\rule{\columnwidth}{0.4pt}}
		\noindent480-Expenditure\\
		
		\columnbreak
		If you lose: \\
		
		\noindent240 (Endowment)\\
		-(Expenditure)\\ \\
		\noindent\makebox[\linewidth]{\rule{\columnwidth}{0.4pt}}
		\noindent240-Expenditure\\
		
	\end{multicols}	
	\subsection*{Your earnings}
	
	You will compete in a series of 25 rounds, and will be paid your payoff for one of them, chosen at random. Prior to the experiment beginning a random number was drawn and placed in an envelope in front of the class. That number contains the round you will be paid for. At the end of the experiment the round number you were paid for and the payout will be displayed for you to confirm the output. Are there any questions at this time?
	
	We will begin with a non paying practice round to familiarize yourself with the controls. For this round only you can input other players' decisions and observe your probability of winning the prize. 

\subsubsection*{Practice}
Before you start making decisions in part 2 of the experiment, we will go over a practice round. These decisions do not affect your earnings and are designed to allow you to become familiar with the interface. In this practice part only, you will be able to enter decisions both for yourself and for other members of your group.

To begin with try entering a value of 50 for each players' expenditure. You will see that the probability of winning is 1/3.

Using an example from the instructions you would enter 100 points for yourself, 20 points for one other member of your group and 40 points for the final group member. The probability that is displayed is 62.5\%

$$ \frac{100}{100+20+40}=\frac{100}{160}=\frac{5}{8}=62.5\%$$

Take a few moments to input values and see how the probability of winning changes. When you are ready to continue press the continue button at the bottom of the screen

\newpage

\section{Regressions with individual characteristics}
\label{app_dem}

\begin{table}[h!]
\centering
{\small 
\begin{tabular}{lcccccccc} \hline\hline
Treatments	& (3) & (1,2) & (1,2) & (2,1) & (2,1) & (1,1,1) & (1,1,1) & (1,1,1) \\
Investment & $x_1$ & $x_1$ & $x_2$ & $x_1$ & $x_2$ & $x_1$ & $x_2$ & $x_3$ \\
	\hline
	Female & -4.457 & -0.497 & 2.492 & 4.988 & 24.55* & -14.76 & 17.22 & 18.41 \\
	& (18.03) & (17.35) & (10.51) & (13.63) & (12.03) & (9.685) & (15.84) & (17.17) \\
	RA & -8.198*** & -4.566 & -0.406 & -6.472 & -0.966 & -13.64** & -1.201 & -11.56*** \\
	& (2.376) & (3.561) & (3.266) & (4.588) & (3.005) & (5.631) & (2.948) & (3.349) \\
	Competitiveness & 1.530 & 3.254 & -9.419 & 8.772 & 4.833 & -24.48 & -13.12** & -1.659 \\
	& (7.602) & (6.315) & (5.452) & (4.948) & (7.199) & (15.86) & (3.979) & (9.353) \\
	Quantitative & -3.936 & -6.304 & -11.87 & 0.718 & -8.749 & 38.94* & 8.375 & 8.731 \\
	& (10.82) & (13.93) & (11.02) & (14.30) & (14.10) & (17.58) & (9.441) & (12.44) \\
	Intercept & 134.4*** & 111.7*** & 99.83*** & 80.03** & 44.80* & 197.3*** & 84.59*** & 135.1*** \\
	& (31.88) & (27.97) & (29.44) & (33.73) & (23.31) & (53.31) & (12.67) & (17.68) \\
	\hline
	Clusters & 9 & 10 & 10 & 9 & 9 & 9 & 9 & 9 \\
	Rounds   & 25 & 25 & 25 & 25 & 25 & 25 & 25 & 25 \\
	$N$ & 2,025 & 750 & 1,500 & 1,350 & 675 & 675 & 675 & 675 \\
	R-squared & 0.043 & 0.012 & 0.019 & 0.036 & 0.032 & 0.327 & 0.064 & 0.090 \\ \hline\hline
\end{tabular}
}
\caption*{Pooled OLS regression results. Robust standard errors in parentheses clustered by matching group. Significance levels: *** $p<$0.01, ** $p<$0.05, * $p<$0.1. $Female$ is a dummy variable equal 1 if the subject self-identified as female, zero otherwise. $RA$ is the number of safe choices in the \cite{HoltLaury} risk aversion elicitation task. $Competitiveness$ is self-reported on a Likert scale between 1 (``much less competitive than average'') and 5 (``much more competitive than average''). $Quantitative$ is a dummy variable based on self-reported university major, equal 1 for majors in the sciences, engineering, mathematics, economics and business, and zero otherwise.}
\end{table}

\newpage

\section{Mathematica code for Section \ref{sec:br}}
\label{app_calc}

Each of the following fragments of code defines linear-quadratic response functions based on the estimates from Table \ref{tab:br_est} and then computes the optimal investment for first movers. In treatments (1,2) and (1,1,1), the computation relies on numerical maximization of the payoff function. In treatment (2,1), it relies on solving a symmetric first-order condition for the equilibrium between two first movers.

\begin{lstlisting}[language=Mathematica]
(* Treatment (1,2) *)
R2[x_]:=0.091*x+0.000096*x^2+62.72
NMaximize[(240+119.73)*x/(x+2*R2[x])-x,{x,0,240}]

(* Treatment (2,1) *)
R2[x_]:=0.249*x-0.0020*x^2+67.60
R2p[x_]:=0.249-0.0040*x
FindRoot[{(240+119.73)*(x+R2[x]-x/2*R2p[x])==(2*x+R2[x])^2},{x,0,240}]

(* Treatment (1,1,1) *)
R2[x_]:=0.103*x-0.00071*x^2+63.93
R3[x1_,x2_]:=0.20*x1-0.00012*x1^2+0.333*x2-0.00081*x2^2+66.76
NMaximize[(240+119.73)*x/(x+R2[x]+R3[x,R2[x]])-x,{x,0,240}]
\end{lstlisting}

\end{document}